\documentclass[preprint,prc,showpacs,twoside,showpacs]{revtex4}
\usepackage{graphicx}
\usepackage{dcolumn}
\usepackage{bm}
\usepackage{color}
\usepackage{amsfonts}
\usepackage{mathrsfs}

\newcommand{\dfrac}{\displaystyle\frac}

\begin{document}
\title{Configuration mixing of angular-momentum projected
triaxial relativistic mean-field wave functions. II. Microscopic analysis of
low-lying states in magnesium isotopes}
\author{J. M. Yao, H. Mei, H. Chen}
\address{School of Physical Science and Technology, Southwest
University, Chongqing 400715, China}
 \author{J. Meng}
 \address{State Key Laboratory of Nuclear Physics and Technology,
  School of Physics, Peking University, Beijing 100871, China}
 \address{School of Physics and Nuclear Energy Engineering, Beihang University, Beijing 100191, China}
 \author{P. Ring}
 \address{Physik-Department der Technischen Universit\"at M\"unchen, D-85748
         Garching, Germany}
\author{D. Vretenar}
 \address{Physics Department, Faculty of Science, University of Zagreb, 10000 Zagreb, Croatia}

\begin{abstract}
The recently developed structure model that uses the generator
coordinate method to perform configuration mixing of
angular-momentum projected wave functions, generated by constrained
self-consistent relativistic mean-field calculations for triaxial
shapes (3DAMP+GCM), is applied in a systematic study of ground
states and low-energy collective states in the  even-even magnesium
isotopes $^{20-40}$Mg. Results obtained using a relativistic
point-coupling nucleon-nucleon effective interaction in the
particle-hole channel, and a density-independent
$\delta$-interaction in the pairing channel, are compared to data
and with previous axial 1DAMP+GCM calculations, both with a
relativistic density functional and the non-relativistic Gogny
force. The effects of the inclusion of triaxial degrees of freedom
on the low-energy spectra and E2 transitions of magnesium isotopes
are examined.
\end{abstract}

 \pacs{21.10.Ky, 21.10.Re, 21.30.Fe, 21.60.Jz, 27.30.+t}
 \maketitle


 \section{Introduction}
 \label{intro}

In the first part of this work \cite{Yao10gcm} the simple mean-field
(single-reference) implementation of the framework of relativistic
energy density functionals (EDF) has been extended to include
long-range correlations related to restoration of symmetries broken
by the static mean field and to fluctuations of collective
coordinates around the mean-field minimum. A model has been
developed that uses the generator coordinate method (GCM) to perform
configuration mixing of three-dimensional angular-momentum projected
(3DAMP) relativistic mean-field wave functions, generated  by
constrained self-consistent calculations for triaxial nuclear
shapes. This calculational framework can be used to perform detailed
studies of low-energy collective excitation spectra and
corresponding electromagnetic transition rates. The particular
implementation of the relativistic 3DAMP+GCM model has been tested
in the calculation of spectroscopic properties of low-spin states in
$^{24}$Mg, in comparison with data and with the results of the
recent work of Ref.~\cite{Bender08}, where a similar 3DAMP+GCM model
has been developed, but based on non-relativistic Skyrme triaxial
mean-field states that are projected both on particle number and
angular momentum, and mixed by the GCM. We note that, very recently,
a new 3DAMP+GCM model with particle number projection has been
implemented, based on the non-relativistic Gogny
force~\cite{Rodriguez10}.

In this work we apply the relativistic 3DAMP+GCM model to a
systematic study of ground states and low-energy collective states
in the  even-even magnesium isotopes $^{20-40}$Mg. The low-energy
structure of magnesium nuclei has attracted considerable interest in
the last decade, both experimental and theoretical. In particular,
the sequence of isotopes $^{20-40}$Mg encompasses three spherical
magic shell numbers:  $N=8$, 20, 28, and, therefore, presents an
excellent case for studies of the evolution of shell structure with
neutron number, weakening of spherical shell closures, disappearance
of magic numbers, and the occurrence of  ``islands of inversion"
\cite{Sorlin08}. Following the pioneering measurement of the
transition rate $B(E2; 0^+_1 \to 2^+_1)$ in the neutron-rich nucleus
$^{32}$Mg \cite{Motobayashi95} that confirmed a large deformation of
this nucleus indicated by the low excitation energy of the $2^+_1$
state \cite{Guillemaud84}, extensive experimental studies of the
low-energy structure of Mg isotopes have been carried out at
RIKEN~\cite{Iwasaki01,Takeuchi09},
 MSU~\cite{Pritychenko99,Cook06,Gade07prl,Gade07prc}, GANIL~\cite{Chiste01}
 and CERN~\cite{Niedermaier05,Schwerdtfeger09}.

In addition to numerous theoretical studies based on large-scale shell-model
calculations \cite{Caurier98,Utsuno99,Otsuka01,Otsuka01a,Caurier05,Marechal05},
the self-consistent mean-field framework, including the non-relativistic Hartree-Fock-Bogolibov
(HFB) model with Skyrme~\cite{Terasaki97} and Gogny forces~\cite{Guzman02}, and
the relativistic mean-field (RMF) model~\cite{Patra91,Ren96},
as well as the macroscopic-microscopic model based on a modified Nilsson potential~\cite{Zhi06},
have been used to analyze the ground-state properties (binding energies, charge radii and deformations)
and low-lying excitation spectra of magnesium isotopes. Of course, to calculate excitation spectra and electromagnetic transition rates it is necessary to go beyond the mean-field approximation and include
dynamic correlations related to the restoration of broken symmetries and to fluctuations of collective
coordinates. Based on the 1DAMP+GCM (axial symmetry) framework, studies of low-energy
spectra of specific Mg isotopes have been performed using non-relativistic models with
 Skyrme~\cite{Valor00,Bender03} and Gogny~\cite{Guzman02} forces,
 as well as relativistic density functionals~\cite{Niksic06I,Niksic06II}.

 In Section~\ref{Sec.II} we present a brief outline of
 the relativistic 3DAMP+GCM model used in the present analysis.
 Section~\ref{Sec.III} describes a study of low-lying collective states of the
 even-even magnesium isotopes $^{20-40}$Mg. A brief summary and an outlook
 for future studies are included in Section~\ref{Sec.IV}.


\section{The 3DAMP+GCM model}
 \label{Sec.II}

 The 1DAMP+GCM calculational framework, restricted to axially symmetric nuclei, has
 recently been extended to include triaxial shapes. 3DAMP+GCM models have been
 developed, based on the self-consistent  Hartree-Fock-Bogoliubov approach with
 Skyrme forces~\cite{Bender08} and the Gogny force~\cite{Rodriguez10}. Starting from
 relativistic energy density functionals, we have implemented a model for
  configuration mixing of three-dimensional angular-momentum
projected (3DAMP) relativistic mean-field wave functions,
generated  by constrained self-consistent calculations
for triaxial nuclear shapes. The details of the model and the numerical tests are
described in Refs.~\cite{Yao09amp,Yao10gcm}. Here we only outline the
basic features of the model that will be used in the study of
 low-lying states in even-even magnesium isotopes $^{20-40}$Mg.

In the 3DAMP+GCM framework the trial angular-momentum projected GCM collective
wave function $\vert\Psi^{JM}_\alpha\rangle$, an eigenfunction of $\hat J^2, \hat J_z$ with
eigenvalue $J(J+1)\hbar^2$ and $M\hbar$, reads
 \begin{eqnarray}
 \label{TrialWF}
 \vert\Psi^{JM}_\alpha\rangle
  &=&  \int d^2q \sum_{K\geq0}f^{JK}_{\alpha}(q)\frac{1}{(1+\delta_{K0})}
  [\hat P^J_{MK}+(-1)^J\hat P^J_{M-K}]\vert\Phi(q)\rangle
 \end{eqnarray}
where $\alpha=1,2,\cdots$ labels collective eigenstates for a given
angular momentum $J$, and $q$ is the generic notation for the
deformation parameters $\beta$ and $\gamma$.  The projection of the
angular momentum $J$ along the intrinsic $z$-axis $K$ takes only
non-negative even values, and $\hat P^J_{MK}$ denotes the
angular-momentum projection operator:
  \begin{eqnarray}
  \hat P^J_{MK}
  =\frac{2J+1}{8\pi^2}\int d\Omega D^{J\ast}_{MK}(\Omega) \hat R(\Omega) \;.
  \end{eqnarray}
 $\Omega$ denotes the set of three Euler angles: \{$\phi, \theta, \psi$\}, and
 $d\Omega=d\phi \sin\theta d\theta d\psi$. $D^J_{MK}(\Omega)$ is the Wigner
 $D$-function, and the rotational operator reads
 $ \hat R(\Omega)=e^{i\phi\hat J_z}e^{i\theta\hat J_y}e^{i\psi\hat J_z}$.
 The set of deformed intrinsic wave functions $\vert \Phi(q)\rangle$ is
 generated by imposing constraints on the axial $q_{20}$ and triaxial $q_{22}$ mass
 quadrupole moments in self-consistent RMF+BCS calculations.

The weight functions $f^{JK}_{\alpha}(q)$ in the collective wave
function Eq.~(\ref{TrialWF}) are obtained from the solution of the
Hill-Wheeler-Griffin (HWG) integral equation:
 \begin{eqnarray}
 \label{HWEq}
 \int dq^\prime\sum_{K^\prime\geq0}
 \left[\mathscr{H}^J_{KK^\prime}(q,q^\prime)
 - E^J_\alpha\mathscr{N}^J_{KK^\prime}(q,q^\prime)\right]
  f^{JK^\prime}_\alpha(q^\prime)=0,
  \end{eqnarray}
 where $\mathscr{H}$ and $\mathscr{N}$ are the angular-momentum projected GCM
 kernel matrices of the Hamiltonian and the norm, respectively~\cite{Yao10gcm}.
 The solution of Eq.~(\ref{HWEq}) determines both the
 energies $E^J_\alpha$ and the amplitudes $f^{JK}_{\alpha}(q)$ of collective states
 $\vert \Psi^{JM}_{\alpha}\rangle$ with good angular momentum.
 The center-of-mass correction to the total energy of the state $J^\pi_\alpha$
 is calculated in the zeroth order of the Kamlah approximation.

 Since the weight functions $f^{JK}_{\alpha}(q)$ are not orthogonal and
 cannot be interpreted as collective wave functions for the deformation
 variables, the collective wave functions $g^J_\alpha(i)$ are determined from
 the eigenstates of norm overlap kernel
 \begin{eqnarray}
 \label{probability}
 g^J_\alpha(i) =\sum\limits_{k}g_k^{J\alpha}u^J_k(i) \;.
 \end{eqnarray}
 These functions are orthonormal and
 \begin{equation}
  \sum_i\vert g^J_\alpha(i)\vert^2=1,
 \end{equation}
 where the sum is over $i\equiv \{K,q\}$.
 The coefficients $g_k^{J\alpha}$ are solutions of the following equation
 \begin{equation}
 \label{CollectiveEQ}
 \sum_{l}\mathcal{H}^{J}_{kl}g_l^{J\alpha}
 = E^J_{\alpha}g_k^{J\alpha}\;,
 \end{equation}
 which is equivalent to Eq.(\ref{HWEq}).
 The matrix $\mathcal{H}^{J}_{kl}$ is determined by the angular-momentum projected GCM
 kernel matrix of the Hamiltonian
 \begin{equation}
 \label{Hcoll}
 \mathcal{H}^{J}_{kl} = \frac{1}{\sqrt{n^J_k}}\frac{1}{\sqrt{n^J_l}}
  \sum_{i,j}{u^J_k(i) \mathscr{H}^J (i,j)u^J_l(j)},
 \end{equation}
 where $n^J_k$ and $u^J_k$ are the non-vanishing
 eigenvalues and eigenvectors
 of the norm overlap kernel $\mathscr{N}^J(i,j)$, respectively.

 The $B(E2)$ value for a transition from an initial state
 $(J_i,\alpha_i)$ to a final state $(J_f,\alpha_f)$ is calculated from
 \begin{eqnarray}
 B(E2;J_i,\alpha_i\rightarrow J_f,\alpha_f)
 =\frac{e^2}{2J_i+1}
        \left|\sum_{q_f,q_i}\langle J_f,q_f\vert\vert \hat Q_{2}\vert\vert J_i,q_i \rangle
        \right|^2 \; ,
 \end{eqnarray}
 where the reduced matrix element is defined by
 \begin{eqnarray}
 \label{Integration2}
 \langle J_f,q_f\vert\vert \hat Q_{2}\vert\vert J_i,q_i\rangle
 =(2J_f+1)
 \sum_{K_iK_f}f^{\ast J_fK_f}_{\alpha_f}(q_f)f^{J_iK_i}_{\alpha_i}(q_i)~~~~~\\
 \times
 \sum_{\mu K^\prime}(-1)^{J_f-K_f}
 \left(\begin{array}{ccc}
   J_f  &  2         &J_i \\
   -K_f & \mu        &K^\prime \\
 \end{array}
 \right) \langle\Phi(q_f)\vert
\hat{Q}^{}_{2\mu} \hat{P}^{J_i}_{K^\prime K_i}\vert\Phi(q_i)\rangle \; ,
\nonumber
 \end{eqnarray}
with $f^{JK}_{\alpha}(q) = (-1)^{J}f^{J-K}_{\alpha}(q)$ for $K<0$.
The spectroscopic quadrupole moment for the state ($J^\pi_\alpha$)
is defined by the expression
 \begin{widetext}\begin{eqnarray}
 Q^{\rm spec}(J^\pi_\alpha)
 &=&e\sqrt{\dfrac{16\pi}{5}}\langle J, M=J,\alpha\vert \hat Q_{20}\vert J,
 M=J,\alpha\rangle\nonumber\\
 &=&e\sqrt{\dfrac{16\pi}{5}}
  \left(\begin{array}{ccc}
   J  &  2         & J \\
   J  &  0         & -J \\
 \end{array} \right)
 \sum_{q_i,q_j} \sum_{KK^\prime}f^{\ast JK^\prime}_{\alpha}(q_j)f^{JK}_{\alpha}(q_i) \nonumber\\
 &&\times
  (2J+1) (-1)^{J+K^\prime}
 \sum_{\mu K^{\prime\prime}}
 \left(\begin{array}{ccc}
   J  &  2         & J \\
   K^{\prime\prime}  &  \mu         & -K^{\prime} \\
 \end{array} \right)
 \langle \Phi(q_j)\vert \hat Q_{2\mu}\hat P^J_{K^{\prime\prime}K}\vert\Phi(q_i)\rangle .
 \end{eqnarray}\end{widetext}
  The matrix elements of the
 charge quadrupole operator $\hat Q_{2\mu}=e\sum_pr^2_pY_{2\mu}(\Omega_p)$
 are calculated in the full configuration space. There is no need for effective charges,
 and $e$ simply corresponds to the bare value of the proton charge.


\section{Low-lying states in magnesium isotopes: results and discussion}
 \label{Sec.III}

As in the first part of this work~\cite{Yao10gcm}, we use the
relativistic point-coupling interaction PC-F1 \cite{Burvenich02} in
the particle-hole channel, and the corresponding
density-independent $\delta$-force in the particle-particle
channel. The parameters of the PC-F1 functional and the pairing
strength constants $V_n$ and $V_p$ have been adjusted simultaneously
to the nuclear matter equation of state, and to ground-state
observables (binding energies, charge and diffraction radii, surface
thickness and pairing gaps) of spherical nuclei \cite{Burvenich02},
with pairing correlations treated in the BCS approximation. In
particular, the pairing strength parameters for neutrons and protons
are $V_n= -308$ MeV fm$^3$ and $V_p= -321$ MeV fm$^3$, respectively.

Parity, $D_{2}$ symmetry, and time-reversal invariance are imposed
in the constrained mean-field calculation of the binding energy map
of an, in general triaxial, even-even nucleus. To solve the Dirac equation for
triaxially deformed potentials, the single-nucleon spinors are expanded in the
basis of eigenfunctions of a three-dimensional harmonic oscillator
(HO) in Cartesian coordinates, with $N_{\mathrm{sh}}=8$ major shells
for $^{20-26}$Mg and $N_{\mathrm{sh}}=10$ for $^{28-40}$Mg. These
numbers of oscillator shells
are sufficient to obtain a reasonably converged mean-field potential
energy surface~\cite{Yao09amp,Yao10gcm}. The HO basis is chosen
isotropic, that is the oscillator parameters $b_{x}=b_{y}=b_{z}=b_{0}
= \sqrt{\hbar/m\omega_{0}}$ in order to keep the basis closed under
rotations~\cite{Egido93,Robledo94}.
 The oscillator frequency is given by $\hbar\omega_{0}=41A^{-1/3}$.
 The Gaussian-Legendre quadrature is used for integrals over
 the Euler angles $\phi, \theta, \psi$ in the calculation of
 the norm and hamiltonian kernels. The choice of the number of mesh points
 for the Euler angles in the interval $[0,\pi]$ is: $N_\phi=N_\psi=8$, and $N_\theta=12$.
 In the 3DAMP+GCM calculations of $^{24}$Mg it has been shown that,
 because of very few level crossings as function of deformation, redundancies
 appear very quickly in the norm kernel when more states are added to the
 nonorthogonal basis~\cite{Bender08,Yao10gcm}. The generator coordinates are,
therefore, chosen in the intervals $0\leq\beta\leq1.2$ and $0\leq\gamma\leq60^\circ$,
with steps $\Delta\beta=0.2$ and $\Delta\gamma=20^\circ$, respectively.
Moreover, eigenstates of the norm overlap kernel with very
small eigenvalues $n^J_k/n^J_{\rm max} < \zeta$ are removed from the
GCM basis. With the cutoff parameter $\zeta=5\times10^{-3}$ for $^{20-26}$Mg, and
$\zeta=1\times10^{-4}$ for $^{28-40}$Mg, fully converged results are obtained
for all low-lying states with $J<6$.

\begin{figure}[h!]
 \centering
 \includegraphics[width=15cm]{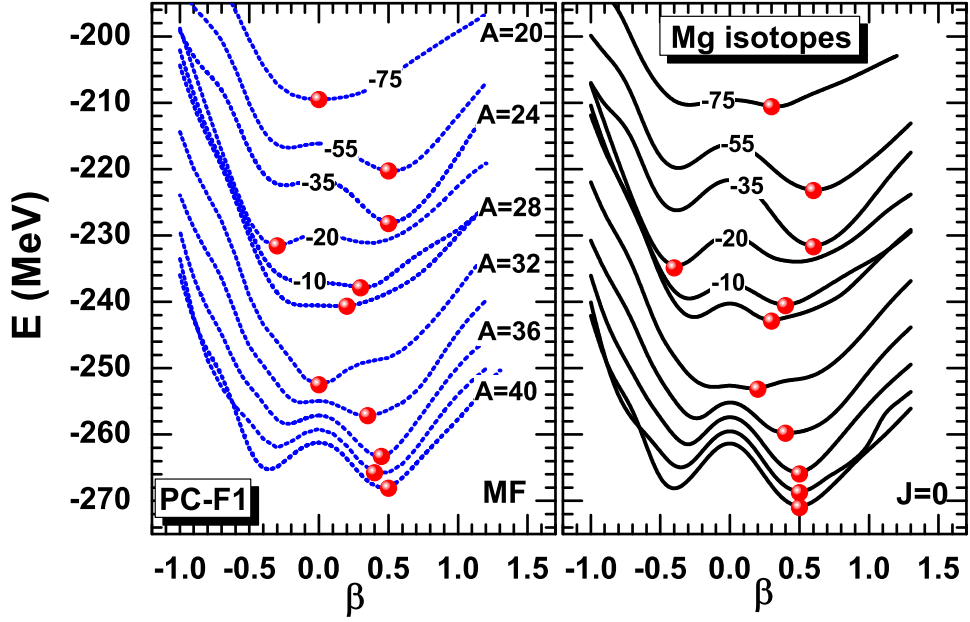}
 \caption{(Color online)  Self-consistent RMF+BCS mean-field (left panel),
 and angular-momentum projected $0^+$ potential energy curves (PEC)
 (right panel) of even-even magnesium isotopes, as functions of the
 axial deformation parameter $\beta$. To plot all the curves in the same
 figure, the PECs of $^{20-28}$Mg have been shifted by
 -75, -55, -35, -20, and -10 MeV, respectively. The position of the
 minimum of each PEC is indicated by a red dot.}
 \label{PES}
 \end{figure}

 In Fig.~\ref{PES} we plot the self-consistent RMF+BCS mean-field,
  and the corresponding angular momentum projected
 ($J^\pi=0^+$), energy curves (PEC) for the even-even
 magnesium isotopes $^{20-40}$Mg, as functions of the axial deformation
  $\beta$ ($\gamma=0$). One might notice an interesting evolution of the
  mean-field PECs from a spherical shape at magic neutron number $N=8$,
 through pronounced prolate shapes, coexistence of oblate and prolate
 shapes, and again to a spherical shape at $N=20$. Increasing further the
 neutron number from $N=20$ to $N=28$, the mean-field minima become
 markedly prolate. Furthermore,
 the effect of angular momentum projection can be inferred from a
 comparison with the corresponding ($J^\pi=0^+$) PECs in the right
 panel of Fig.~\ref{PES}. In particular, in the neighborhood of the spherical minimum
 the $J=0$ PECs of $^{20,32}$Mg are very soft with respect to $\beta$. In other
 isotopes the deformed minima become deeper after projection.

\begin{figure}[h!]
 \centering
 \includegraphics[width=12cm]{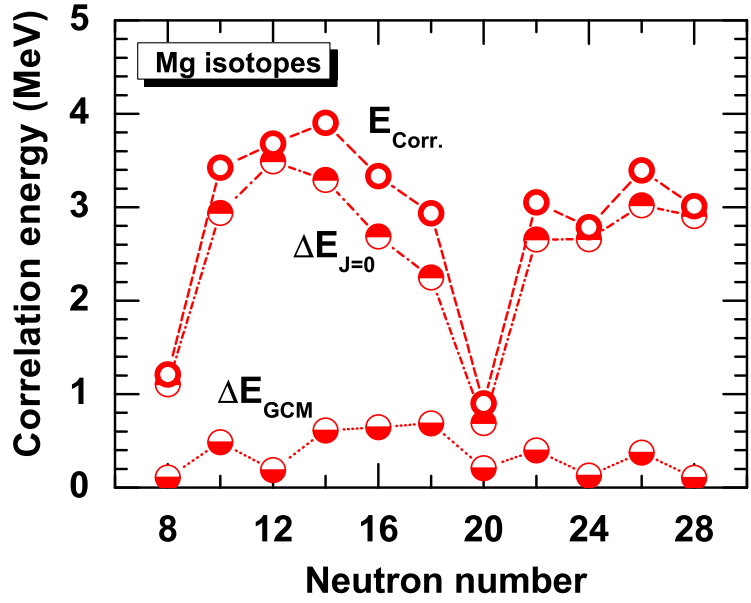} 
 \caption{(Color online) Total ground-state dynamical correlation energies
 $E_{\rm Corr}$ of Mg  isotopes, as a function of the number of neutrons.
 $E_{\rm Corr}$ is the sum of the rotational energy correction
  $\Delta E_{J=0}$ and the energy gained by configuration mixing
  $\Delta E_{\rm GCM}$.}
 \label{correlation}
 \end{figure}

Figure~\ref{correlation} displays the total ground-state dynamical correlation energies of Mg
 isotopes, as a function of the number of neutrons. As shown in the figure,  $E_{\rm Corr}$
 consists of a rotational energy correction $\Delta E_{J=0}$ that results from the restoration of
 rotational symmetry
 \begin{equation}
  \Delta E_{J=0}     =  E_{J=0}(\beta_0)-E_{\rm MF}(\beta_m)\;,
 \end{equation}
 and the correlation energy gained by GCM configuration mixing
  \begin{equation}
 \Delta E_{\rm GCM} = E(0^+_1) - E_{J=0}(\beta_0)\; .
 \end{equation}
 $\beta_m$ and $\beta_0$ denote the axial deformation parameters at the minima
 of the mean-field and the ($J^\pi=0^+$) angular-momentum projected PECs,
 respectively (cf. Fig.~\ref{PES}). $E_{\rm Corr}$ shows a strong dependence on
 shape and shell structure. It is large for deformed mid-shell nuclei, with a
 maximum of $\sim 4$ MeV at  $N=14$, and is drastically reduced ($\sim1$~MeV)
 for the two isotopes with the neutron magic numbers $N=8$ and $N=20$.
 Projection on angular momentum $J=0$ , that is, the rotational energy correction
 $\Delta E_{J=0}$ constitutes the dominant part of the total dynamical correlation energy.
 This is generally valid for a great majority of nuclei, as it has been shown in the
 global study of quadrupole correlation effects \cite{Bender06}, performed with
 GCM configuration mixing of axially symmetric Skyrme-Hartree-Fock+BCS states,
 with the two-point topological Gaussian overlap approximation for angular-momentum
 projection. As also shown in Ref.~\cite{Bender06}, in Fig.~\ref{correlation} one notices
 that the correlation energy $\Delta E_{\rm GCM}$ gained from configuration mixing of
 different deformed states is of the order of several hundreds keV, and not very sensitive
 to nuclear shape and shell structure. $\Delta E_{\rm GCM}$ is, in fact, composed of
 two parts: a potential term that is
 negative and in size comparable to the correlation energy induced by
 angular momentum projection, and a kinetic part (energy
 of the zero-point vibrational motion) that is positive and cancels
 to a large extent the potential term~\cite{Ring80}.

\begin{figure}[h!]
 \centering
 \includegraphics[width=12cm]{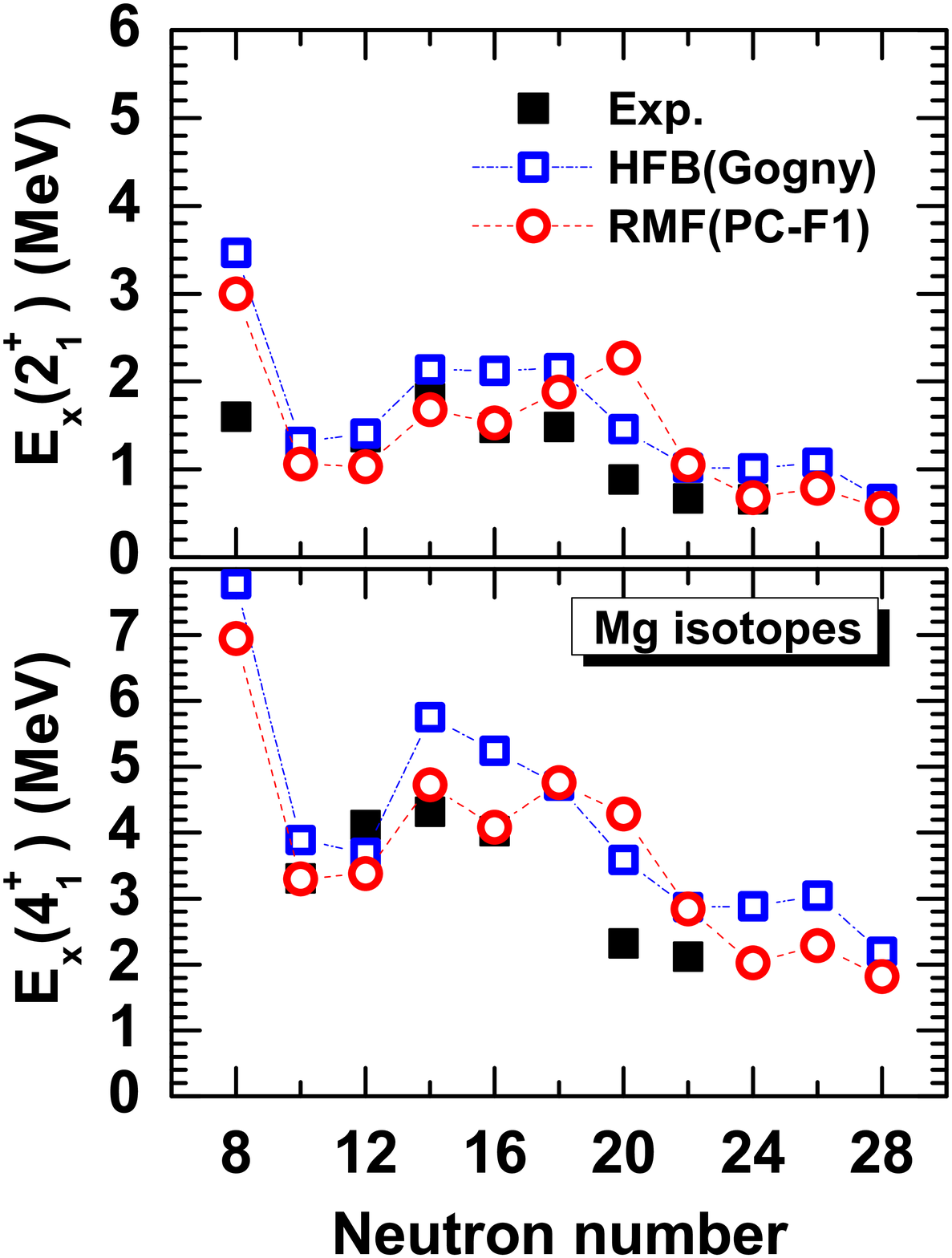} 
 \caption{(Color online) Excitation energies of the states $2^+_1$ and $4^+_1$ in $^{20-40}$Mg,
 calculated using the 1DAMP+GCM model with the relativistic density functional PC-F1,
are compared to available data \cite{Raman01,Gade07prc,Gade07prl} and
 the results of the 1DAMP+GCM calculation based on the
 non-relativistic HFB framework with the Gogny force~\cite{Guzman02}.}
 \label{excitation}
 \end{figure}

The excitation energies of the states $2^+_1$ and $4^+_1$ in $^{20-40}$Mg, calculated
using the 1DAMP+GCM model with the relativistic density functional PC-F1,
are compared in Fig.~\ref{excitation} to the available data and
 the prediction of the 1DAMP+GCM calculation based on the
 non-relativistic HFB framework with the Gogny force~\cite{Guzman02}. Both models
 yield excitation energies of the $2^+_1$ and $4^+_1$ states in reasonable agreement
 with data and, on the average, the values obtained with PC-F1 are $10-30\%$ lower
 than those calculated with the Gogny interaction D1S (except for $^{32}$Mg).
 This is due to relatively weak
 neutron pairing correlations in the present calculation, that lead to an increase of the
 corresponding moment of inertia for the yrast states. As noted in our previous study
 of $^{24}$Mg in Ref.~\cite{Yao10gcm}, the excitation energies of yrast states increase
 when the pairing strength parameters $V_{n/p}$ are adjusted to the pairing gaps
 determined from empirical odd-even mass differences in this particular mass region.
 Both calculations preserve the $N=8$ magic number and with PC-F1 also at $N=20$
 a pronounced shell closure is obtained, whereas the model based on the Gogny
 force predicts a much lower excitation energy of the $2^+_1$ state in $^{32}$Mg,
 in better agreement with data. One might notice, however, that both models
 predict the  $4^+_1$ state in this nucleus at energies far above the experimental
 value. The $N=28$ shell closure disappears in both calculations, and $^{40}$Mg
 is predicted to be prolate deformed.

\begin{figure}[h!]
 \centering
 \includegraphics[width=14cm]{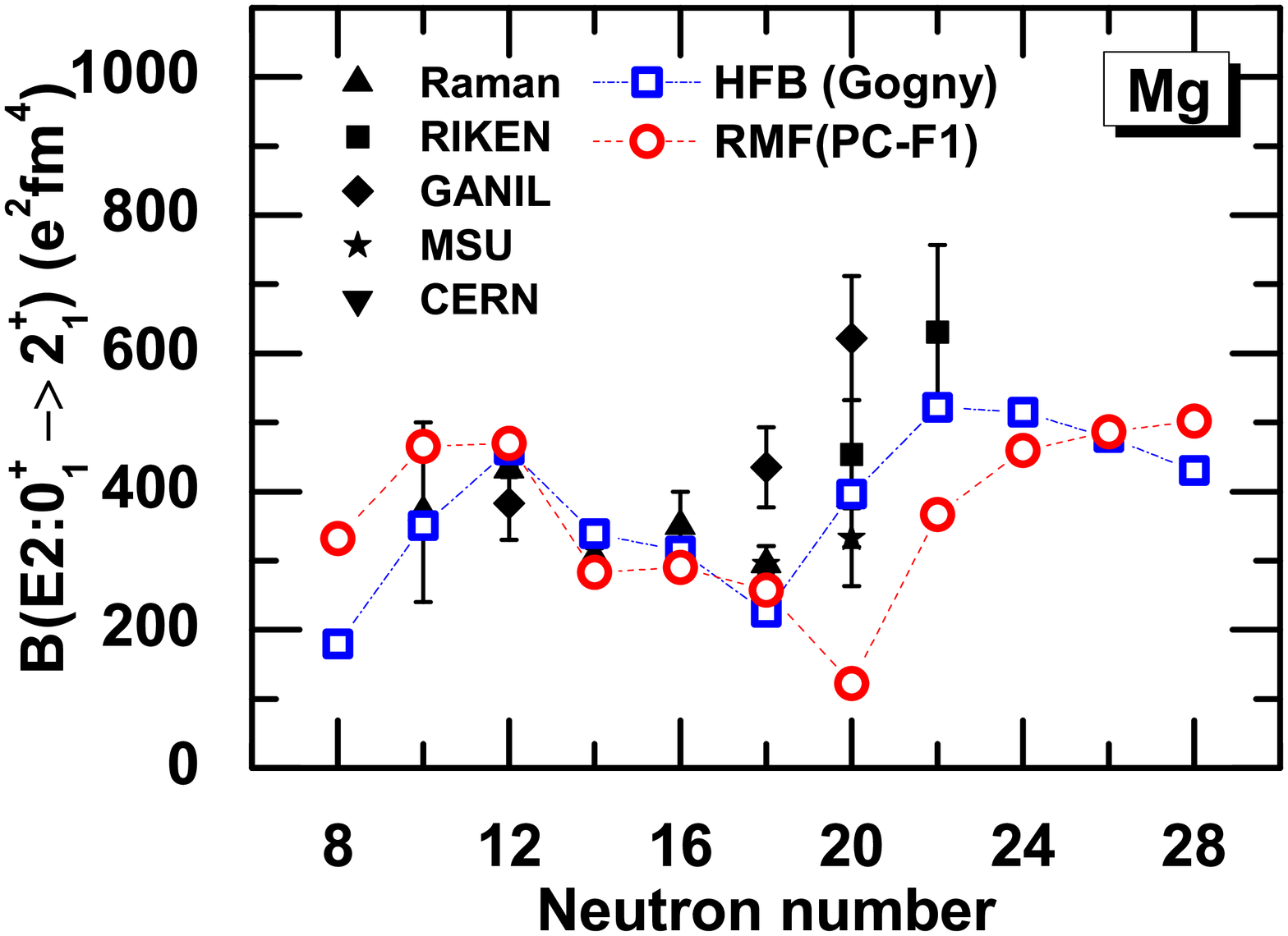} 
 \caption{(Color online) $B(E2; 0^+_1\rightarrow2^+_1)$ (e$^2$fm$^4$) values
 in $^{20-40}$Mg,  calculated using the 1DAMP+GCM model with the relativistic density
 functional PC-F1, are compared to available data
 \cite{Motobayashi95,Pritychenko99,Iwasaki01,Chiste01,Raman01,Niedermaier05} and
 the results of the 1DAMP+GCM calculation based on the
 non-relativistic HFB framework with the Gogny force~\cite{Guzman02}.}
 \label{BE2}
 \end{figure}

The corresponding  $B(E2; 0^+_1\rightarrow2^+_1)$ (e$^2$fm$^4$) values
 in $^{20-40}$Mg are shown in Fig.~\ref{BE2}. 1DAMP+GCM calculations, both
 the present one using the functional PC-F1 and that based on the Gogny force~\cite{Guzman02},
 yield results in reasonable agreement with data except, of course, PC-F1 at and in the
 neighborhood of the neutron number $N=20$. Since the Gogny force D1S predicts an
 axially deformed ground state for $^{32}$Mg, the corresponding B(E2) value for the
 transition $0^+_1\to 2^+_1$ is much closer to the experimental value, compared to
 the calculation with PC-F1 which yields a spherical ground state at  $N=20$.
 The functional PC-F1, together with the density-independent $\delta$-force
 ($V_n= -308$ MeV fm$^3$ and $V_p= -321$ MeV fm$^3$) predicts indeed
 a very small $B(E2)$ value for this transition in $^{32}$Mg.
 In Ref.~\cite{Yao09amp} it has been suggested that a better adjustment of
 pairing strength parameters and eventually the inclusion of triaxiality,
 that is the $\gamma$ degree of freedom, could improve the results for
 $^{32}$Mg. Already in the 1DAMP+GCM axial calculations we have verified
 that, by adjusting the pairing strengths specifically to the empirical pairing gaps
 around $^{32}$Mg (five-point formula):
 $V_n= -465$ MeV fm$^3$ and $V_p= -350$ MeV fm$^3$, the calculated
 transition rate increases to $B(E2; 0^+_1\rightarrow2^+_1) = 313.5$ e$^2$fm$^4$.
 To have a consistent model, however,  in the remaining calculations of this work we
 will continue using the original pairing strengths that were adjusted simultaneously with
 the parameters of the PC-F1 effective interaction in the particle-hole channel \cite{Burvenich02}.

\begin{figure}[h!]
 \centering
 \includegraphics[width=12cm]{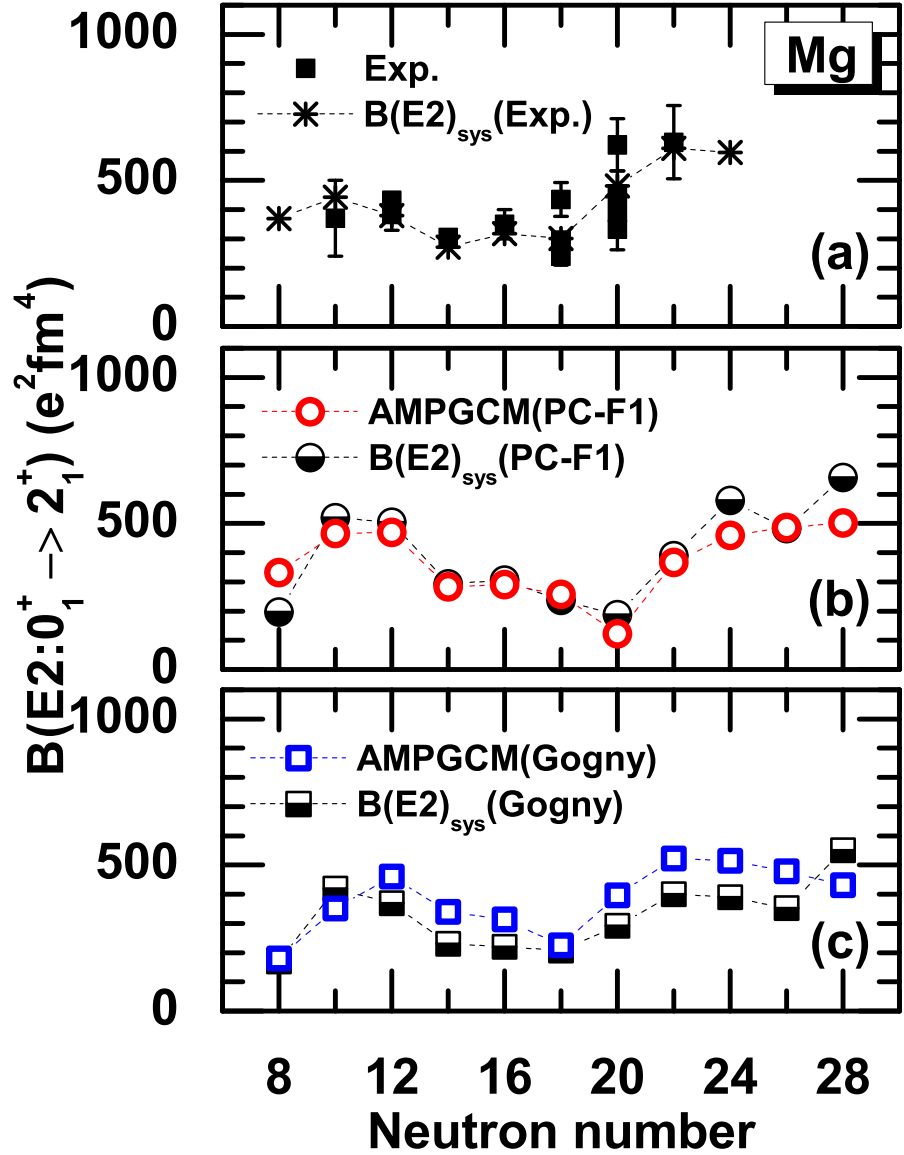} 
 \caption{(Color online)
$B(E2)$ values calculated with the LDM formula
Eq.~(\ref{LDM}) for the transition $0^+_1\rightarrow2^+_1$ in Mg isotopes,
are compared to data  \cite{Motobayashi95,Pritychenko99,Iwasaki01,Chiste01,Raman01,Niedermaier05}
in panel (a), and to the results of the 1DAMP+GCM calculations
with the functional PC-F1 in (b), and with the Gogny force D1S \cite{Guzman02} in (c).}
 \label{BE2_LDM}
 \end{figure}

A measure of collectivity of the lowest excited states in magnesium isotopes can be obtained
by comparing the experimental and calculated $B(E2; 0^+_1\rightarrow2^+_1)$ with the
prediction of an empirical formula based on the liquid-drop model (LDM)~\cite{Raman88},
  \begin{equation}
  \label{LDM}
  B(E2: 0^+_1\rightarrow2^+_1)_{\rm sys.} = 6.47Z^2A^{-0.69}E^{-1}_x(2^+_1)\; .
  \end{equation}
This comparison is shown in Fig.~\ref{BE2_LDM}. In the upper panel the B(E2) values
calculated with the LDM formula are compared to data, whereas in the two lower panels
they are compared to the results of the 1DAMP+GCM calculations
with the functional PC-F1 and with the Gogny force D1S.
The excitation energies $E(2^+_1)$  (in MeV) that appear in the
LDM expression Eq.~(\ref{LDM}), correspond to the experimental values and those
calculated with PC-F1 and Gogny D1S, respectively. One notices a very good
agreement between data and the B(E2) values predicted by the LDM formula. Based
on the recently measured $E(2^+_1)$ values for $^{20}$Mg: $1598(10)$ keV, and $^{36}$Mg:
$660(6)$ keV,  Eq.~(\ref{LDM}) predicts the corresponding $B(E2; 0^+_1\rightarrow2^+_1)$
values 368.9(23) e$^2$fm$^4$ and 595.4(54) e$^2$fm$^4$, respectively.
The 1DAMP+GCM  calculation based on the PC-F1 functional yields somewhat smaller
B(E2) values for the  $0^+_1\rightarrow2^+_1$ transition in $^{20}$Mg (332 e$^2$fm$^4$)
and $^{36}$Mg (460 e$^2$fm$^4$).

\begin{figure}[h!]
 \centering
 \includegraphics[width=12cm]{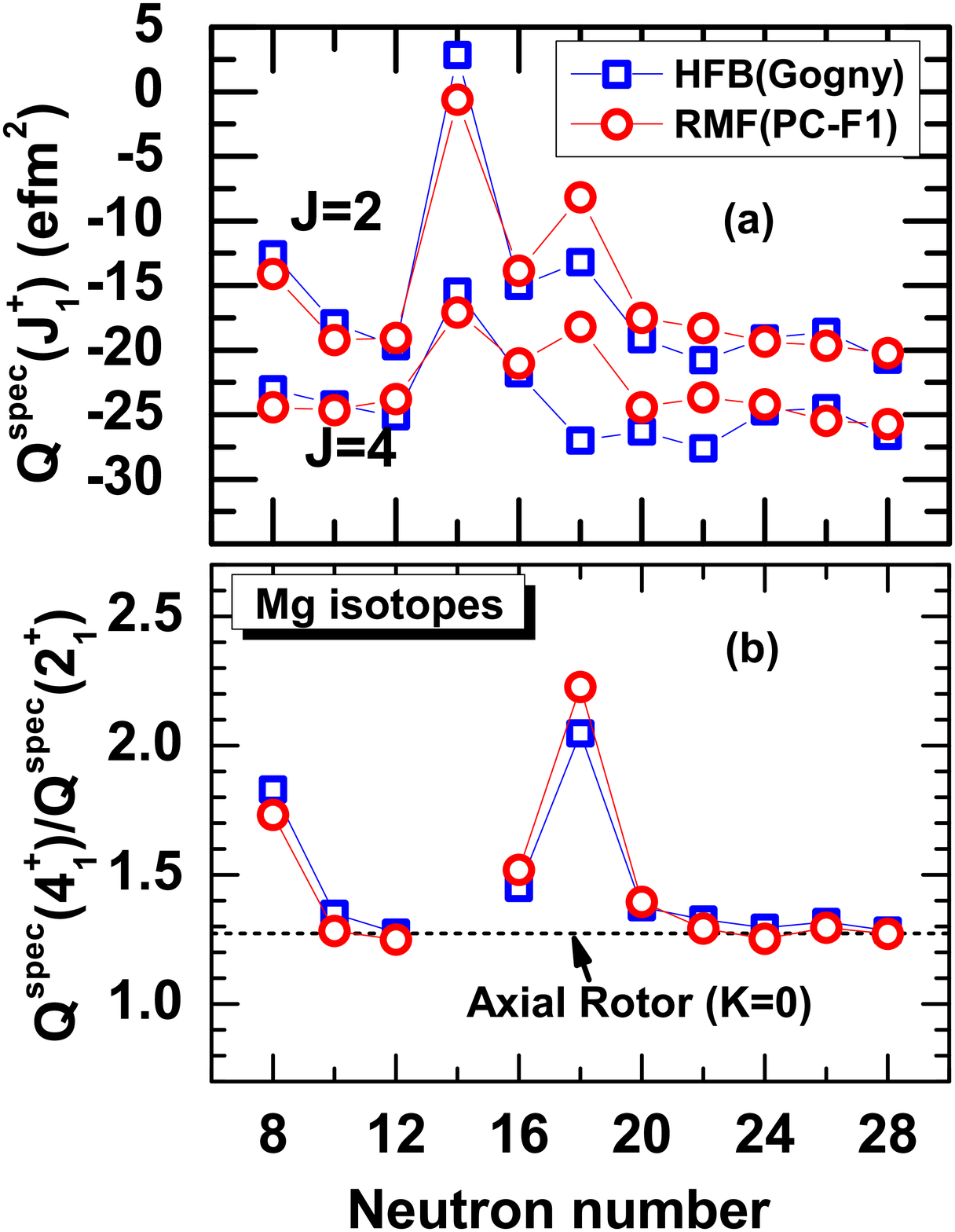} 
 \caption{(Color online) Spectroscopic quadrupole moments
of the states $2^+_1$ and $4^+_1$ in $^{20-40}$Mg, calculated
using the 1DAMP+GCM model with the relativistic density functional PC-F1,
and the corresponding values based on the
non-relativistic HFB framework with the Gogny force~\cite{Guzman02}
(upper panel). The calculated ratios $Q^{\rm spec}(4^+_1)/Q^{\rm spec}(2^+_1)$
are compared to the value that corresponds to a rigid axial rotor with $K=0$
(lower panel).}
 \label{spectroscopic}
 \end{figure}

In Fig.~\ref{spectroscopic} we plot the spectroscopic quadrupole moments
of the states $2^+_1$ and $4^+_1$ in $^{20-40}$Mg, calculated
using the 1DAMP+GCM model with the relativistic density functional PC-F1,
and compared to the corresponding values based on the
non-relativistic HFB framework with the Gogny force~\cite{Guzman02}.
One might notice a very good agreement between the results of the two
model calculations, with the exception of $^{30}$Mg. In the lower panel
the calculated ratios $Q^{\rm spec}(4^+_1)/Q^{\rm spec}(2^+_1)$ are compared to
 the value that corresponds to a rigid axial rotor with $K=0$, that is $\approx 1.27$ .
 In $^{26}$Mg both models predict a very small value of
 $Q^{\rm spec}(2^+_1)$, and this gives rise to an exceptionally high value
 of $Q^{\rm spec}(4^+_1)/Q^{\rm spec}(2^+_1)$ that does not fit the scale of
 the vertical axis. This result  indicates that there is a large contribution from
 nonzero-$K$ components in the yrast band of $^{26}$Mg. Large deviations
 from the axial rotor value are also predicted for $^{20}$Mg and $^{30}$Mg.
 For the isotopes $^{22,24,28,32-40}$Mg both models yield
 $Q^{\rm spec}(4^+_1)/Q^{\rm spec}(2^+_1)$ quite close to that of rigid axial
 rotor. Note that this is also true in $^{32}$Mg, for which the calculation based
 on the Gogny force yields a deformed ground state, whereas this state is
 spherical in the present axially symmetric calculation using the functional
PC-F1. In both calculations, however, the states $2^+_1$ and $4^+_1$
are prolate deformed.

\begin{figure}[h!]
 \centering
 \includegraphics[width=12cm]{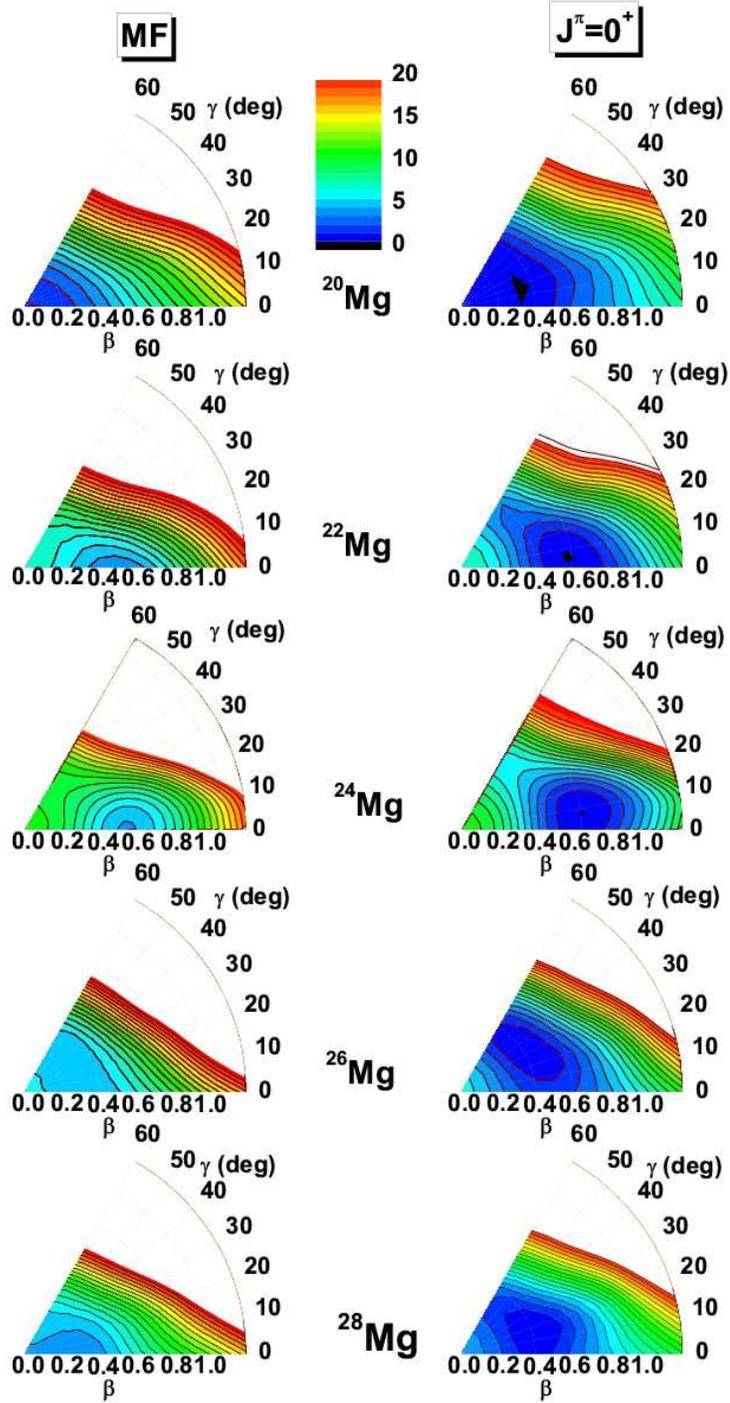} 
 \caption{(Color online)
 Self-consistent RMF+BCS triaxial quadrupole
binding energy maps of the even-even $^{20-28}$Mg isotopes
in the $\beta - \gamma$ plane ($0\le \gamma\le 60^0$) (left panel),
and the corresponding angular-momentum $J^\pi=0^+$
projected energy surfaces (right panel).
All energies are normalized with respect to
the binding energy of the absolute minimum, the contours join points
on the surface with the same energy (in MeV).}
 \label{PESs1}
 \end{figure}

\begin{figure}[h!]
 \centering
 \includegraphics[width=11cm]{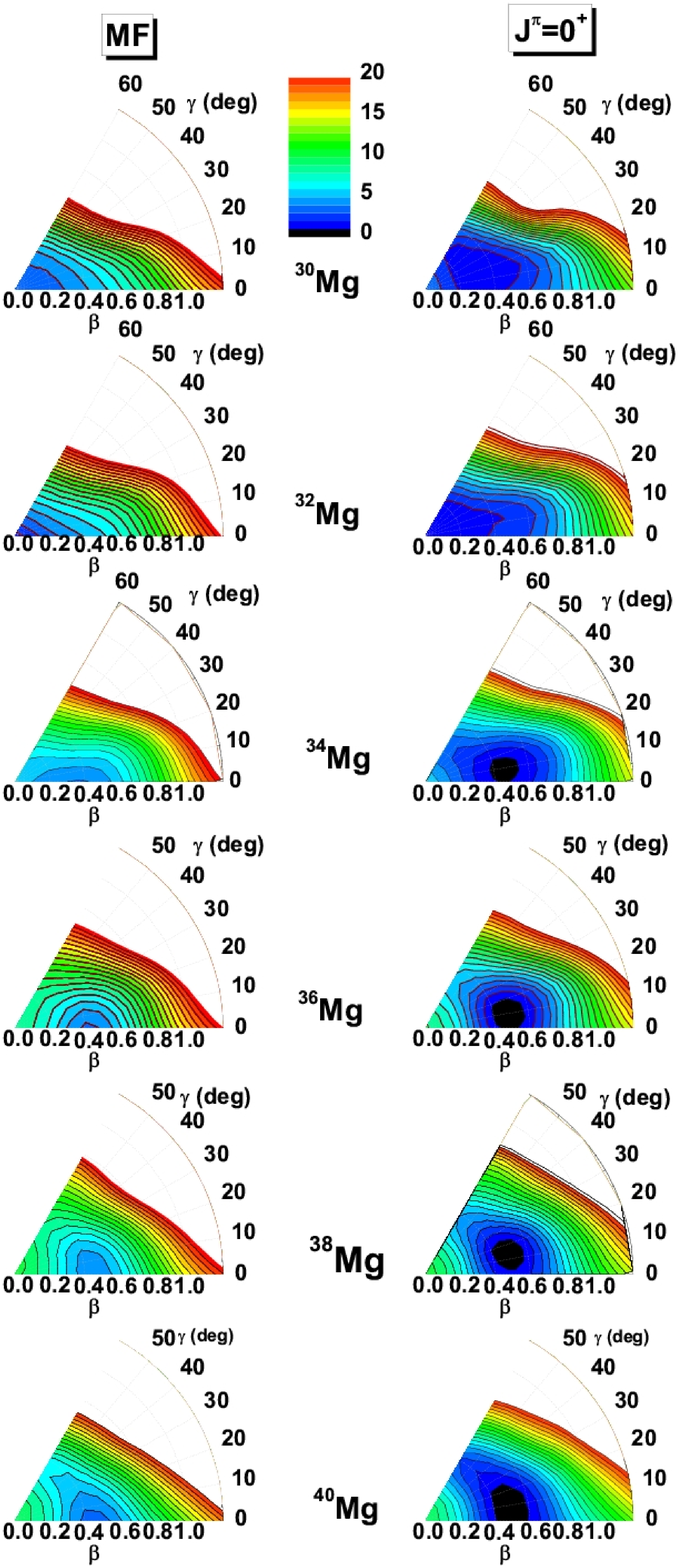} 
 \caption{(Color online) Same as described in the caption to Fig.~\ref{PESs1} but for the isotopes
  $^{30-40}$Mg.}
 \label{PESs2}
 \end{figure}

 To examine the influence of triaxiality, that is, of including the
 $\gamma$ degree of freedom on the spectroscopic properties of low-lying states in
 magnesium isotopes, we have performed full 3DAMP+GCM calculations using the
 relativistic functional PC-F1. In Figs.~\ref{PESs1} and \ref{PESs2} we display the
 resulting self-consistent RMF+BCS triaxial quadrupole
binding energy maps of the even-even $^{20-40}$Mg isotopes
in the $\beta - \gamma$ plane ($0\le \gamma\le 60^0$),
and the corresponding angular-momentum $J^\pi=0^+$
projected energy surfaces. All energies are normalized with respect to
the binding energy of the absolute minimum, the contours join points
on the surface with the same energy (in MeV). In general the inclusion
of the triaxial deformation degree of freedom reduces considerably the
barriers separating axially prolate and oblate minima in the well-deformed
isotopes $^{22,24,34-40}$Mg. We also notice that the angular-momentum $J^\pi=0^+$
projected energy surfaces of  $^{26-32}$Mg are rather soft both in $\beta$ and $\gamma$.

\begin{figure}[h!]
 \centering
 \includegraphics[width=16 cm]{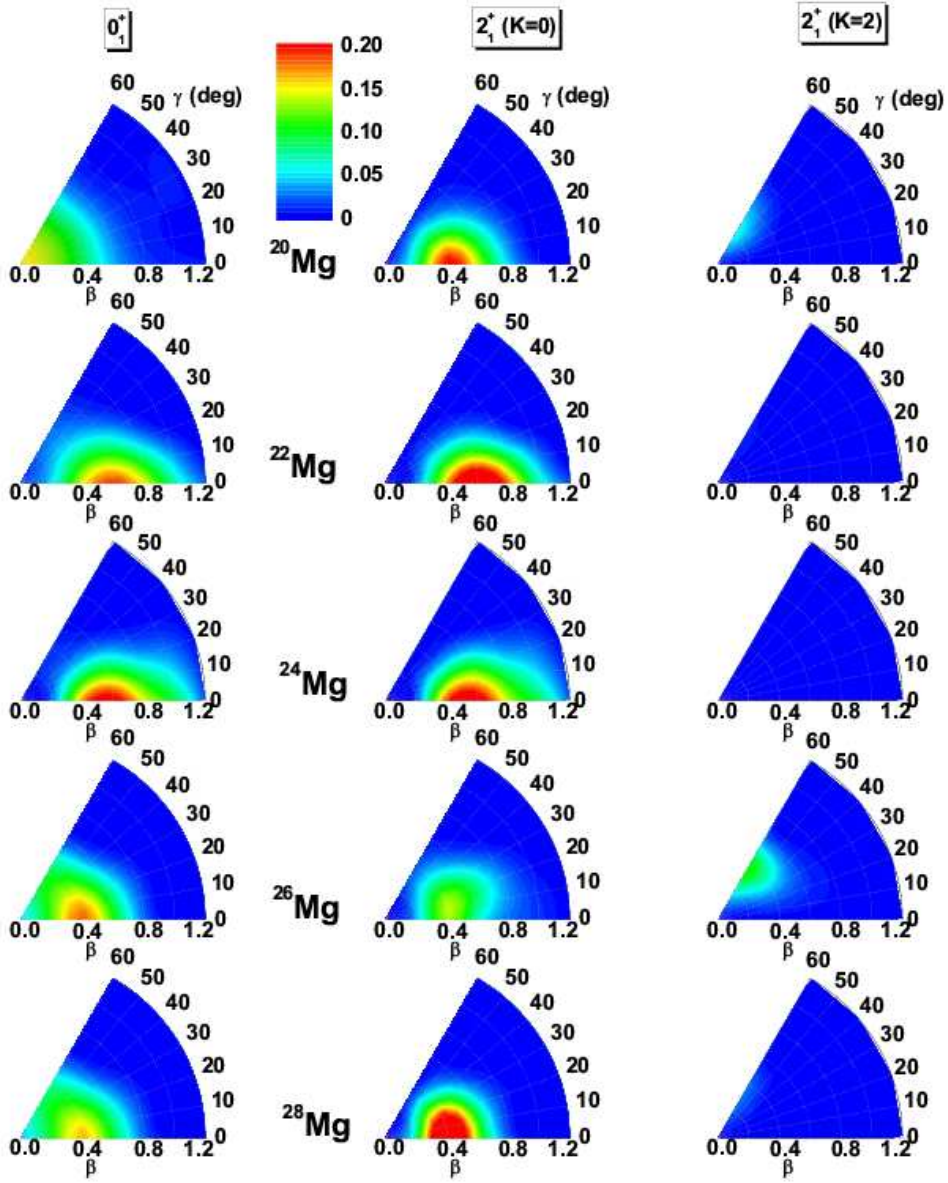} 
 \caption{(Color online) Probability
distributions $\vert g^J_\alpha\vert^2$  of the collective wave functions Eq.~(\ref{probability})
in the $\beta - \gamma$ plane, for the  the states of $0^+_1$ and
 $2^+_1$ (both the $K=0$ and $K=2$ components) of  $^{20-28}$Mg. }
 \label{WFs1}
 \end{figure}

\begin{figure}[h!]
 \centering
 \includegraphics[width=16cm]{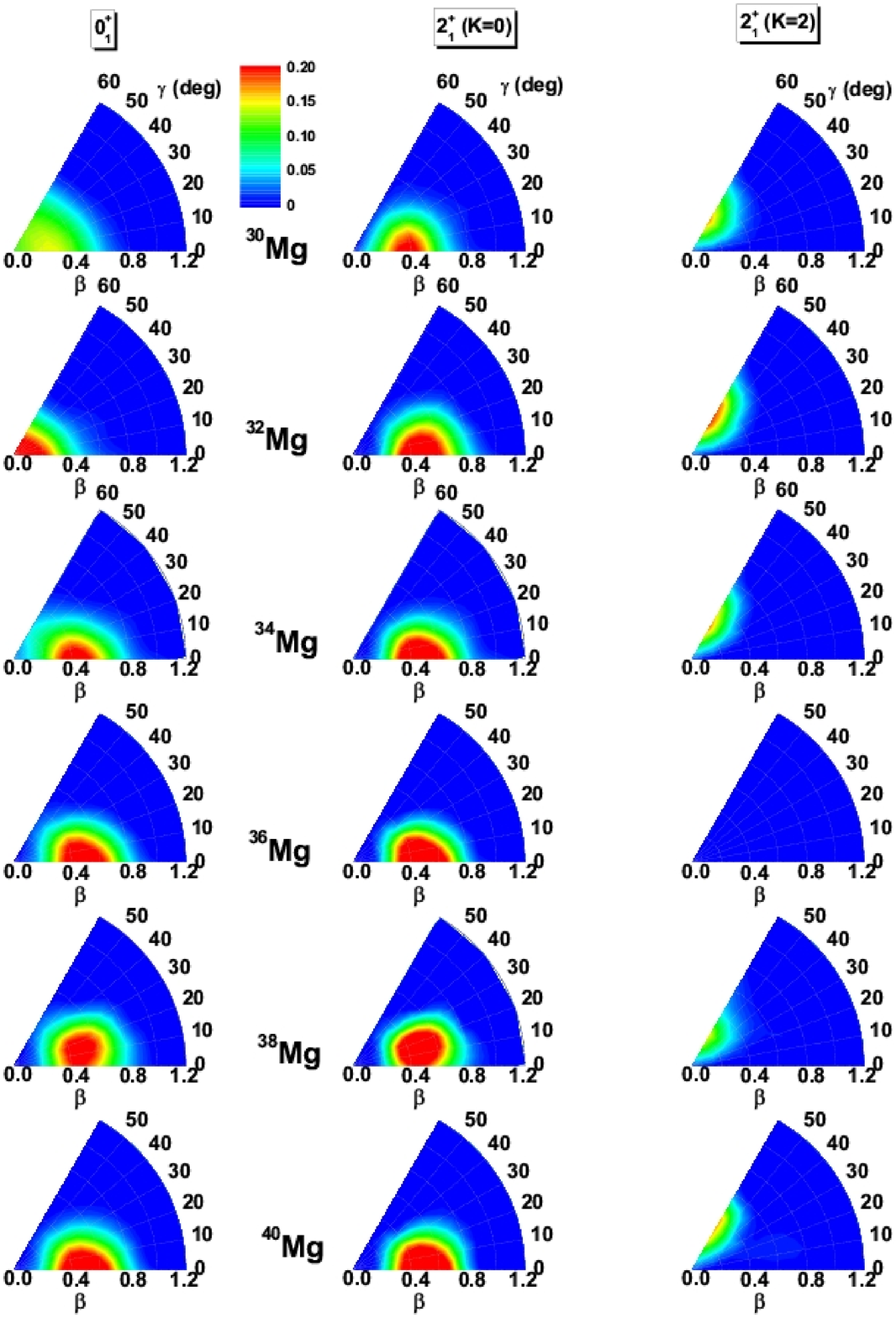} 
 \caption{(Color online) Same as described in the caption to Fig.~\ref{WFs1} but for the isotopes
 $^{30-40}$Mg. }
 \label{WFs2}
 \end{figure}

\begin{figure}[h!]
 \centering
 \includegraphics[width=14cm]{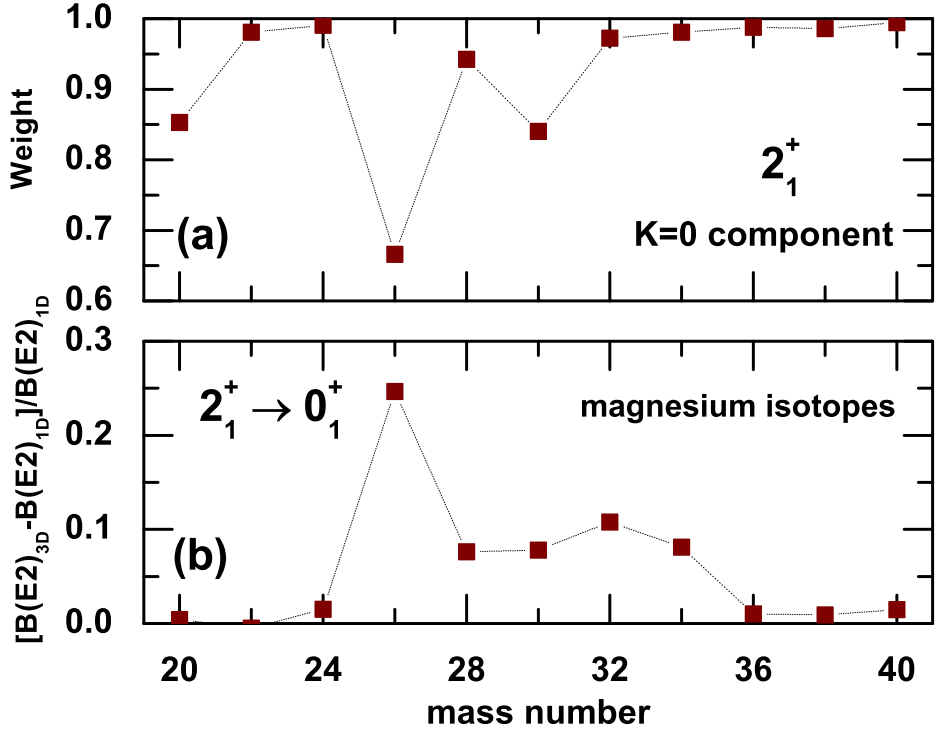} 
 \caption{(Color online) Upper panel: relative weight of the $K=0$ component in
 the collective wave functions of the $2^+_1$ states of magnesium isotopes $^{20-40}$Mg.
 Lower panel: differences between the
  $B(E2;2^+_1\rightarrow0^+_1)$ values calculated in the 3DAMP+GCM and the
  1DAMP+GCM models, normalized to the 1D values. }
 \label{Weights}
 \end{figure}

The low-energy excitation spectra and collective wave functions are calculated as solutions
of the Hill-Wheller-Griffin integral equation for each angular momentum, and thus take
into account fluctuations of the collective coordinates $\beta$ and $\gamma$ around the mean-field minima.
For the sequence of isotopes $^{20-40}$Mg Figs.~\ref{WFs1} and \ref{WFs2} display the probability
distributions $\vert g^J_\alpha\vert^2$  of the collective wave functions Eq.~(\ref{probability})
in the $\beta - \gamma$ plane, for the states $0^+_1$ and
 $2^+_1$ (both the $K=0$ and $K=2$ components). It appears that $^{20,30,32}$Mg are spherical
in the ground state, whereas all the other isotopes are prolate deformed and the ground-state
deformation is especially pronounced in heavier Mg nuclei. The first excited state $2^+_1$ is
prolate deformed in all Mg nuclei, even in $^{32}$Mg. In several isotopes, most notably in
$^{26}$Mg and $^{30}$Mg, the collective wave function of the state $2^+_1$ contains
sizeable admixtures of the $K=2$ component. This can be seen more clearly in Fig.~\ref{Weights}
where, after integrating the probability distributions over $\beta$ and
 $\gamma$, in the upper panel we plot the relative weight of the $K=0$ component in
 the collective wave functions of the $2^+_1$ states of magnesium isotopes $^{20-40}$Mg.
 The softness toward triaxial shapes is especially pronounced in $^{20}$Mg, $^{26}$Mg,
 and $^{30}$Mg. The contribution of the $K=2$ component in the wave functions of $2^+_1$
 will generally affect the calculated B(E2) values for transitions to the ground state.
 In the lower panel of Fig.~\ref{Weights} we show the differences between the
  $B(E2;2^+_1\rightarrow0^+_1)$ values calculated in the full 3DAMP+GCM and the
  axial 1DAMP+GCM models, normalized to the 1D values. A marked effect of $K$-mixing is
  found not only in $^{26}$Mg, but also in some heavier isotopes including $^{32}$Mg.

 Finally, a quantitative comparison between the axial 1DAMP + GCM and the
full 3DAMP + GCM calculations for $^{20-40}$Mg, based on the relativistic functional
 PC-F1, is presented in Table~\ref{tab1}. The ground-state energies $E_{\rm gs}$ (in MeV),
 excitation energies of the $2^+_1$ and $4^+_1$ states (in MeV), and
  $B(E2\downarrow; J\rightarrow J-2)$ values (in e$^2$fm$^4$)
   for the lowest states with $J=2^+,4^+$ in magnesium isotopes are included in the
   table. In general the inclusion of the $\gamma$ degree of freedom leads to the
   lowering of the binding energies of low-lying states and to an increase of the
   calculated B(E2) values. The latter is particularly prominent in $^{26}$Mg, in
   which the 3DAMP + GCM yields an enhancement of $\approx 25\%$ for the
   $B(E2;0^+_1\rightarrow 2^+_1)$. Especially interesting is the case of  $^{32}$Mg, which
   shows a pronounced lowering of the excitation energies of  $2^+_1$ and $4^+_1$,
  whereas the binding energy of the ground-state, being spherical, is not influenced
  by the inclusion of triaxial shapes. These excitation energies are, however, still far
  above the experimental energies and even though the
   $B(E2; 0^+_1\rightarrow2^+_1)$ value increases by $\approx 10\%$, it is about a
   factor three smaller than the empirical value. However,  when the pairing strength
   parameters are adjusted specifically to the empirical pairing gaps
 around $^{32}$Mg (five-point formula):
 $V_n= -465$ MeV fm$^3$ and $V_p= -350$ MeV fm$^3$, the calculated
 transition rate increases to $B(E2; 0^+_1\rightarrow2^+_1) = 330.1$ e$^2$fm$^4$, in
 rather good agreement with data. This results shows the importance of a more detailed
 study of pairing correlations in $N \approx 20$ neutron-rich nuclei.


  \section{Summary}
 \label{Sec.IV}
 The very successful framework of relativistic energy density functionals has
 mostly been used on the mean-field level to describe ground-state properties
 of medium-heavy and heavy nuclei. When considering applications, however,
 it is important to develop EDF-based structure models that go beyond the static
mean-field approximation. Detailed predictions of excitation spectra and transition
rates necessitate the inclusion of correlations related to the restoration of
broken symmetries and to fluctuations of collective variables. In recent years
several new models have been developed that extend the relativistic EDF-based
approach and perform the restoration of symmetries broken by the static
mean field and take into account fluctuations around the mean-field minimum.
This is relatively simple in the case of axial symmetry, that is, when only
one collective coordinate is considered \cite{Niksic06I,Niksic06II}, but such
models become much more involved, technically complicated, and computationally
demanding when possible triaxial shapes are taken into account.

In Refs.~\cite{Yao09amp} and \cite{Yao10gcm} we have implemented and
tested a new model that uses the generator coordinate method (GCM)
to perform configuration mixing of three-dimensional angular-momentum
projected (3DAMP) relativistic mean-field wave functions,
generated  by constrained self-consistent calculations
for triaxial nuclear shapes. In the present study this calculational framework
has been used to analyze the influence of triaxiality on the
low-energy collective excitation spectra and the corresponding electric
quadrupole transition rates of even-even magnesium isotopes $^{20-40}$Mg.
The self-consistent solutions of the constrained RMF+BCS equations have
been obtained using the relativistic point-coupling interaction PC-F1 \cite{Burvenich02}
in the particle-hole channel, and a density-independent $\delta$-force
in the particle-particle channel. Since the low-energy spectra of $^{20-40}$Mg
were previously investigated in the axial 1DAMP+GCM model based
based on the non-relativistic HFB framework with the Gogny force \cite{Guzman02},
in the first instance we have performed axial 1D calculations and compared the
results with data and those obtained in Ref.~\cite{Guzman02}. In general, a
good agreement has been obtained between the results of the two model
calculations, except for $^{30,32}$Mg.  The low excitation energy of $2^+_1$
and the large $B(E2; 0^+_1 \to 2^+_1)$ indicate that the neutron rich nucleus $^{32}$Mg
is deformed, even though the number of neutrons equal the ``spherical magic number"
$N=20$. The data are reproduced reasonably well by the 1DAMP+GCM model based
on the Gogny force, which yields a deformed ground state for $^{32}$Mg. The
present axial calculation, on the other hand, predicts a spherical $\beta$-soft
ground state for $^{32}$Mg, although the lowest excited states
$2^+_1$ and $4^+_1$ are calculated to be prolate deformed. The corresponding
$B(E2; 0^+_1 \to 2^+_1)$ is much smaller than the experimental value. Both models
predict prolate ground states for heavier Mg isotopes, including the $N=28$ nucleus
$^{40}$Mg.

To analyze the effect of triaxiality and K-mixing on the low-energy structure of
Mg isotopes, we have also performed a full 3DAMP+GCM calculation based
on the relativistic density functional PC-F1 and a  density-independent
$\delta$ pairing interaction. When compared with the 1DAMP+GCM results, it is noted
that the inclusion of the $\gamma$ degree of freedom leads to the lowering of the
binding energies of low-lying states and to an increase of the calculated B(E2) values
in deformed isotopes. In several isotopes a pronounced degree of $\gamma$ softness
and K-mixing is predicted for the yrast states. The effect is strongest in $^{26}$Mg, in
which the 3DAMP + GCM yields an enhancement of $\approx 25\%$ for the
$B(E2;0^+_1\rightarrow 2^+_1)$. Even in the triaxial case the functional PC-F1 preserves
the spherical shell closure at $N=20$, i.e., it predicts a spherical ground state for
$^{32}$Mg. The excitation energies of the states $2^+_1$ and $4^+_1$ in this
nucleus are lowered considerably with respect to the axial case, but they are still
much higher than the experimental values. Correspondingly, the calculated
$B(E2; 0^+_1\rightarrow2^+_1)$ is about a  factor three smaller than the empirical value.
It is noted, however, that when the pairing strength parameters are adjusted specifically to
the empirical pairing gaps around $^{32}$Mg, the calculated
transition rate increases to $B(E2; 0^+_1\rightarrow2^+_1) = 330.1$ e$^2$fm$^4$, much
closer to the available data.

In future studies the 3DAMP + GCM model based on relativistic density functionals will be
applied to the description of shape transitions and shape coexistence phenomena in
medium-heavy and heavy nuclei. We also plan to compare the results of full 3D
angular-momentum projection and GCM configuration mixing, with those obtained in
the recently developed model for the solution of the
eigenvalue problem of a five-dimensional collective Hamiltonian for
quadrupole vibrational and rotational degrees of freedom, with parameters
determined by constrained self-consistent relativistic mean-field calculations
for triaxial shapes \cite{Niksic09}.


 \section*{Acknowledgments}

This work was partly supported by the Major State 973 Program
2007CB815000 and the NSFC under Grant Nos. 10947013, 10975008 and
10775004, the Southwest University Initial Research Foundation Grant
to Doctor (No. SWU109011), the DFG cluster of excellence
\textquotedblleft Origin and Structure of the
Universe\textquotedblright\ (www.universe-cluster.de), by MZOS -
project 1191005-1010, and by the Chinese-Croatian project "Nuclear
structure far from stability".


\newpage
  \begin{table*}[]
   \centering
   \caption{The ground-state energy $E_{\rm gs}$ (in MeV), excitation energies of the
   $2^+_1$ and $4^+_1$ states (in MeV), and
  $B(E2\uparrow; J-2\rightarrow J)$ values (in e$^2$fm$^4$)
   for the lowest states with $J=2^+,4^+$ in magnesium isotopes. Results obtained
   in the axial 1DAMP+GCM calculation are compared with those of the full
    3DAMP+GCM model.   }
   \begin{tabular}{|c|cccrr|cccrr|}
    \hline\hline
     &   \multicolumn{5}{c|}{1DAMP+GCM}                  &   \multicolumn{5}{c|}{3DAMP+GCM}            \\
    \hline
    Isotopes  & $E_{\rm gs}$ &  $E_x(2^+_1)$  & $E_x(4^+_1)$  &$E2\uparrow(2^+_1)$ & $E2\uparrow(4^+_1)$  & $E_{\rm gs}$ & $E_x(2^+_1)$  &  $E_x(4^+_1)$&$E2\uparrow(2^+_1)$ & $E2\uparrow(4^+_1)$  \\
    \hline
    $^{20}$Mg &-135.501 & 2.999  & 6.948  &332   & 205  & -135.469 & 2.945 & 6.798 & 333 &344  \\
    $^{22}$Mg &-168.246 & 1.063  & 3.298  &465   & 242  & -168.277 & 1.048 & 3.313 & 463 &429\\
    $^{24}$Mg &-196.822 & 1.058  & 3.438  &470   & 233  & -197.064 & 0.927 & 3.203 & 477 &422\\
    $^{26}$Mg &-215.322 & 1.679  & 4.725  &283   & 151  & -215.737 & 1.569 & 4.541 & 353 &355 \\
    $^{28}$Mg &-231.242 & 1.527  & 4.080  &291   & 167  & -231.445 & 1.331 & 3.819 & 313 &319 \\
    $^{30}$Mg &-243.563 & 1.882  & 4.760  &257   & 154  & -243.637 & 1.721 & 4.416 & 277 &313 \\
    $^{32}$Mg &-253.381 & 2.270  & 4.283  &122   & 212  & -253.390 & 1.907 & 3.844 & 136 &413 \\
    $^{34}$Mg &-260.198 & 1.050  & 2.842  &367   & 214  & -260.375 & 0.920 & 2.612 & 397 &419 \\
    $^{36}$Mg &-266.045 & 0.679  & 2.024  &460   & 238  & -266.477 & 0.673 & 2.112 & 465 &430  \\
    $^{38}$Mg &-269.022 & 0.785  & 2.286  &487   & 261  & -269.974 & 0.628 & 2.010 & 491 &456\\
    $^{40}$Mg &-271.098 & 0.556  & 1.815  &502   & 261  & -271.442 & 0.533 & 1.836 & 509 &484 \\
   \hline \hline
 \end{tabular}
 \label{tab1}
 \end{table*}

\newpage

\newpage

\newpage

\newpage

\newpage

\newpage

\newpage

\end{document}